\documentclass[letterpaper, 10 pt, conference]{ieeeconf}  
\IEEEoverridecommandlockouts                              
\usepackage{url}
\usepackage[utf8]{inputenc}
\usepackage[T1]{fontenc}
\usepackage{cite}
\usepackage{amsmath,amssymb,amsfonts}
\usepackage{algorithmic}
\usepackage{graphicx}
\usepackage{textcomp}
\usepackage{xcolor}
\def\BibTeX{{\rm B\kern-.05em{\sc i\kern-.025em b}\kern-.08em
    T\kern-.1667em\lower.7ex\hbox{E}\kern-.125emX}}

\title{\LARGE \bf
Monoscopic vs. Stereoscopic Views and Display Types in the Teleoperation of Unmanned Ground Vehicles for Object Avoidance
}

\author{Yiming Luo$^{1}$, Jialin Wang$^{1}$, Hai-Ning Liang$^{1,*}$, Shan Luo$^{2}$, and Eng Gee Lim$^{3}$
\thanks{The work has been supported in part by Xi'an Jiaotong-Liverpool University (XJTLU) Key Special Fund (KSF-A-03; KSF-P-02) and XJTLU Research Development Fund.}
\thanks{$^{1}$Yiming Luo, Jialin Wang, Hai-Ning Liang are with the Department of Computing, Xi'an Jiaotong-Liverpool University, Suzhou, China.}%
\thanks{$^{2}$Shan Luo is with the Department of Computer Science, The University of Liverpool, UK.}%
\thanks{$^{3}$Eng Geen Lim is with The Department of Communications and Networking, Xi'an Jiaotong-Liverpool University, Suzhou, China.}%
\thanks{$^{*}$Corresponding author ({\tt\small haining.liang@xjtlu.edu.cn}).}%
}

\begin{document}

\maketitle
\thispagestyle{empty}
\pagestyle{empty}

\begin{abstract}
Virtual reality (VR) head-mounted displays (HMD) have recently been used to provide an immersive, first-person vision/view in real-time for manipulating remotely-controlled unmanned ground vehicles (UGV). The teleoperation of UGV can be  challenging for operators when it is done in real time. One big challenge is for operators to perceive quickly and rapidly the distance of objects that are around the UGV while it is moving. In this research, we explore the use of monoscopic and stereoscopic views and display types (immersive and non-immersive VR) for operating vehicles remotely. We conducted two user studies to explore their feasibility and advantages. Results show a significantly better performance when using an immersive display with stereoscopic view for dynamic, real-time navigation tasks that require avoiding both moving and static obstacles. The use of stereoscopic view in an immersive display in particular improved user performance and led to better usability. 
\end{abstract}

\begin{figure*}[t]
\parbox{1\textwidth}{
\centering
\includegraphics[width=1\textwidth]{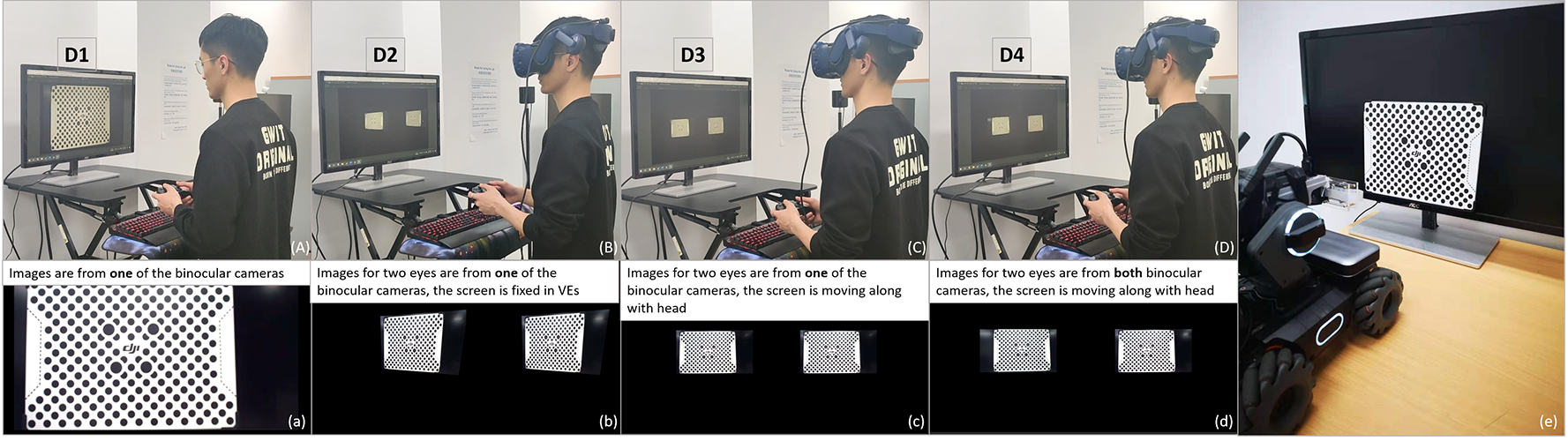}}
\caption{The four display modes tested in the first experiment. (A, a) \textit{D1}: a regular monitor; using one of the binocular cameras’ images. (B, b) \textit{D2}: a fixed virtual screen in VR; using one of the binocular cameras’ images. (C, c) \textit{D3}: a virtual screen in VR moving along with head motions; using one of the binocular cameras’ images (monoscopic). (D, d) \textit{D4}: a virtual screen in VR moving along with head motions; using both of the binocular cameras’ images (stereoscopic).(e) The picture of UGV with the camera facing a board on the monitor.}
\label{fig1}
\end{figure*}

\section{Introduction}
The teleoperation of unmanned ground vehicles (UGV) or unmanned aerial vehicles (UAV) is challenging for operators because it is not easy to have a good understanding of the terrain and surrounding objects given that they are away from the actual environment \cite{b1}. With advances in real-time image transmission, issues such as low image quality, long delays in image transmission, and unstable signal transmission have been improved significantly, making their teleoperation stable and practical in a wide range of situations \cite{b29}. These improvements have allowed for the exploration of different image views that can be shown on different display types to improve the ability of operators to control UGV in a precise and efficient manner.

Current displays used in the teleoperation of UGV include primarily traditional 2D screens (such as those on mobile phones/tablets and typical desktop computers). More recently immersive displays, such as virtual reality head-mounted displays (VR HMD), have started to make their way for UGV teleoperations. VR HMD allow a higher level of immersion and presence (that is, the feeling of being “there” in the environment) \cite{b36}, which can enhance performance \cite{b37,b8}. The disadvantage of VR HMD is that they have small displays and their use may increase the chance of operators having motion sickness, a common issue with HMD that is not present in normal displays \cite{b6,b7}. 

Whether it is a traditional screen or a VR HMD, it is not easy for operators to gauge the distance and depth perception of objects from the video images shown in the displays in real time because, unlike being in the real environment, they often do not provide enough visual stereoscopic information for the operators to see.

Depending on the placement of the camera in the drone, the operator can see images based on the first-person view (FPV) or the third-person view (TPV) \cite{b1,b30,b31}. Mainstream drones in the market can supply a FPV perspective primarily. Based on cost factors and the stability of image transmission, most drones use a single camera, typically of high-quality, to transmit monoscopic images to a 2D non-immersive screen display. In the absence of stereoscopic cues, these 2D images do not often provide enough visual information for operators to perceive clearly and accurately distance and depth of obstacles surrounding the drone\cite{b1,b32}. This situation could be worse when there are moving objects while the drone is moving towards them.

Using binocular cameras could provide additional information to assist operators gauge distance information of objects. The use of binocular cameras has been studied on UAV mainly but not so much on UGV \cite{b1,b32,b33,b34}. At low altitudes but high speed, a UAV with an on-board computer that can provide stereoscopic FPV rendered in a VR HMD could enhance the control of the flying drone \cite{b1}. In our research, we explore the use of binocular cameras in UGV, which tend to go slower than UAV and could have more objects on its path, and whether they can enhance teleporters’ stereoscopic perception and their performance in obstacle avoidance tasks in real-time. In addition, we want to compare the usability and performance in both normal displays and immersive VR HMD.

In this paper, we first review related work about viewing modes and display types as well as obstacle avoidance tasks and their performance metrics. We then present two user studies. The first study explores four viewing modes in a normal 2D display and VR HMD (see Fig.~\ref{fig1}). We compared these modes based on participants' subjective feedback and objective results in a pre-designed obstacle avoidance experiment. The version with stereoscopic view in VR had best performance but participants still faced challenges with moving objects. These results led us modify this viewing mode and run a second study comparing the new mode against it. The results show that the new fifth mode could lead to an improved overall performance. The results of these two studies can inform the design of future teleoperator-UGV interaction that offers better performance and user experience.

\section{Related Work}
In this section, we first introduce previous work on viewing modes and display types for UGV/UAV. Then we present the tasks for testing robot performance and the metrics for their evaluation.

\subsection{Viewing Modes and Display Types}
Viewing modes (e.g., 2D vs. 3D) and display types (VR vs. 2D displays) can affect users' levels of immersion, flow, and performance \cite{b37,b38,b39}. Most teleoperated unmanned systems use a monocular camera that displays video streams captured from the camera attached to a robot. The camera supplies images for operators to see and make decisions for where to go. Because the camera tends to have a small field-of-view (FOV), the video images could only provide limited visual details which in turn can lead to lower performance on tasks such as target detection and identification \cite{b3}. It requires operators to put extra effort to survey the environment (e.g., by manipulating and/or rotating the robot to get different views) \cite{b2}, an inefficient process that can increase mental and visual load and the feeling of motion sickness. Also, important distance cues may not be provided while depth perception may be degraded when the FOV is restricted \cite{b4}. 

When driving a UGV, operators have more difficulties in judging the speed of the vehicle, time-to-collision, perception of objects, locations of obstacles, and the start of a sharp curve \cite{b5}. The level of difficulty can increase if the objects around the drone are moving and not static. However, simply increasing the FOV may lead to other issues. For example, in a typical 2D display, this could mean that the users may need to scan a wide view of the environment, requiring moving their head left and right frequently. In addition, a larger FOV may cause additional usability issues with VR HMD as rotation head movements might induce greater motion sickness \cite{b6,b7}. The challenge is to provide additional visual cues without significantly changing the FOV and causing large head movements, especially rotational ones.

3D Stereoscopic (3DS) views have been found to be better than traditional non-stereo (2D) views on manipulation tasks with either virtual or real objects \cite{b8,b9}. In \cite{b1} the authors explored immersive 3DS views for flying drones which enable higher accurate depth perception and led to better teleoperation and navigation performance. Stereoscopic FPV presents significant advantages over monocular FPV\cite{b13,b14}. The distance between the binocular cameras used to achieve 3DS also has an important effect. The best performance was achieved when the inter-camera distance was less than the inter-ocular distance, which is 2–3 cm and 6 cm, respectively \cite{b15}. However, artificially induced binocular stereo-vision may increase motion sickness and perceived stress \cite{b10}. Latency in image transmission is also associated with motion sickness \cite{b11,b12}. Low image quality caused by reduced frames per second (fps), reduced resolution of the display (pixels per frame), while a lower gray-scale (number of levels of brightness or bits per frame) \cite{b20} can also increase the motion sickness. While motion sickness can be an issue and when it is possible to have high-resolution images without transmission delays \cite{b2}, the depth information and stereoscopic perception provided by the binocular cameras can significantly improve the operator’s teleoperation performance for obstacle avoidance and precise maneuverability.

Distance underestimation and overestimation can occur when objects are viewed in normal displays and especially in immersive VR environments (VE) \cite{b15,b16}. In experiments with teleoperated UGV, it has been found that operators underestimated the distances from obstacles and landmarks \cite{b17}. As such, given the benefits of 3DS views in VR, this research aims to explore whether the combination of 3DS and VR could support operators to gauge distance information during teleoperation manipulation and improve their performance in obstacle avoidance tasks.

\subsection{Tasks and Metrics}
Teleoperation tasks that require real-time obstacle avoidance have been researched in the context of human-drone interaction. For example, in \cite{b18,b19}, researchers explored a remotely controlled robot that has been integrated with a laser sensor and a monocular camera to capture distance information. This combination provided operators with images displayed in a virtual UI. The operator was asked to drive this robot through a series of mazes and avoid obstacles. Its performance was measured using a set of metrics \cite{b28}: (a) Obstacle encounter, the number of collisions of the robot against obstacles; (b) Efficiency, the time to complete tasks; and (c) Subjective ratings, usability issues with controls and the interface. We followed a similar approach and used these metrics in this research.

In addition, it has been shown that moving objects require a greater depth information of the objects to avoid colliding with them \cite{b21}. As such, the maze used in our research (see Fig.~\ref{fig3} on the next page) incorporated both static and dynamic objects of different types. Their combination allows us to explore in detail the comparative performance of teleoperating a UGV based on different viewing modes and display types.  

\section{User Study 1}
To explore how monoscopic and stereoscopic views and display types affect the distance perception of users, we conducted a 4-condition, 8-person within-subjects study. The main task consisted of participants driving a UGV in real time using a game controller through a maze that had both static and moving obstacles. 

\begin{figure}[t]
\parbox{3.4in}{
\centering
\includegraphics[width=3.4in]{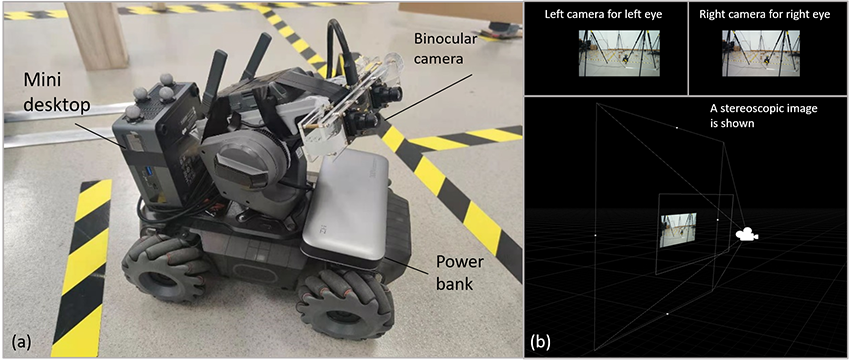}}
\caption{(a) The prototype used in this research. It contains a UGV (DJI RoboMaster S1) and a FPV system which was built with a binocular camera, an on-board mini desktop, and a power bank. (b) Implementation of 3DS view.}
\label{fig2}
\end{figure}

\begin{figure*}[htbp]
\parbox{1\textwidth}{
\centering
\includegraphics[width=1\textwidth]{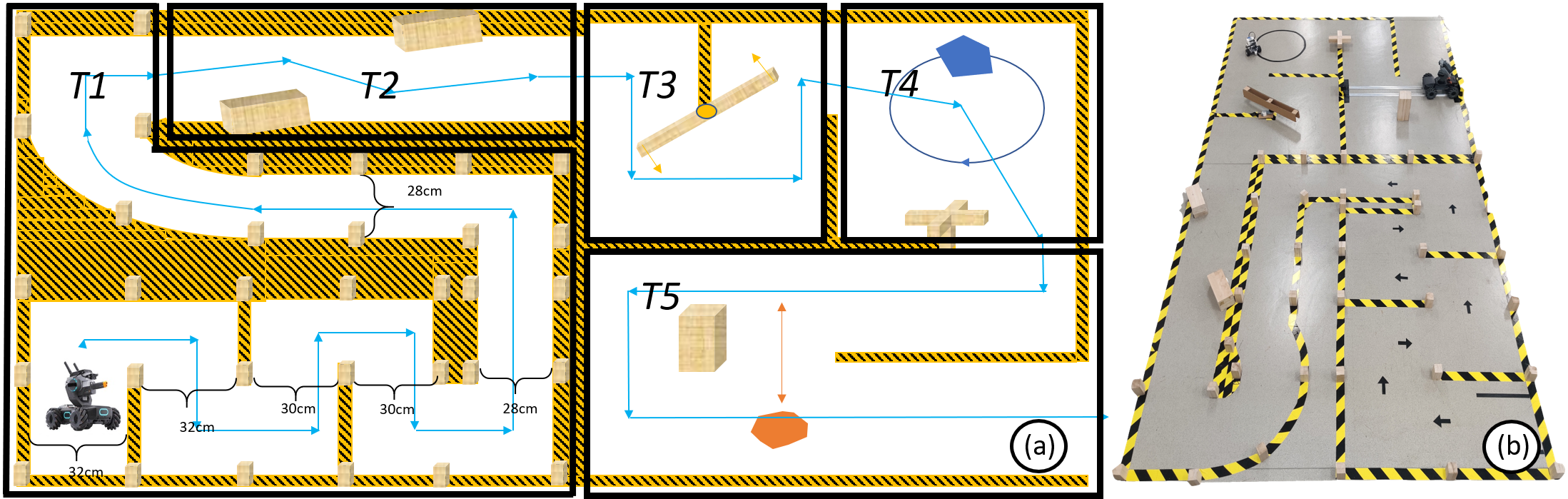}}
\caption{Overview of the maze. (a) The distribution of the different tasks in the maze (\textit{T1}, \textit{T2}, \textit{T3}, \textit{T4}, and \textit{T5}); (b) A picture of the actual maze used in the study.}
\label{fig3}
\end{figure*}

\begin{figure}[htbp]
\parbox{3.4in}{
\centering
\includegraphics[width=3.4in]{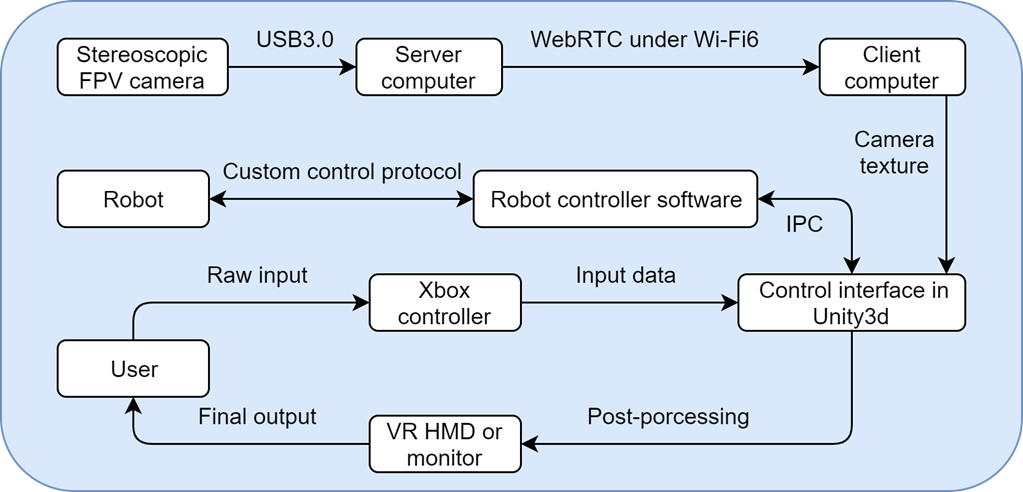}}
\caption{Overview of the components of the control and image transmission system.}
\label{fig4}
\end{figure}

\subsection{Prototype}\label{AA}
Fig.~\ref{fig2}a shows the prototype we developed to conduct this research. A DJI RoboMaster s1\footnote{DJI RoboMaster S1: \url{https://www.dji.com/robomaster-s1}} was used as the mobile UGV. The transmission system consisted of a binocular camera, a mini desktop, and a power bank. The first-person view (FPV) images from the camera were transmitted to a PC and rendered in a traditional 27-inch 4K monitor or VR HMD, which in our case was a HTC Vive Pro Eye\footnote{HTC Vive Pro:\url{ https://www.vive.com/uk/product/vive-pro-eye/specs/}}.

The binocular camera consists of 2 monocular cameras with a resolution of 1280 × 720, a 71° FOV, and  distortion less than ± 0.3 \% \cite{b26}.  The binocular camera has a stereoscopic view with a 2560 × 720 resolution. It was installed on a mounting platform with Inter Pupillary Distance (IPD) adjustment with a range of about 26mm to 84mm. The average human IPD is 65mm, with a range of 54 to 75 millimeter for young adults, between 16 to 24 years \cite{b25}. Therefore, the IPD was set to 65mm in the stereoscopic FPV condition. The right camera was moved to a central position as the input source of the other monoscopic conditions. The FPV system used WebRTC as the video streaming protocol which has a 214 ± 7 (ms) latency at 2560 x 720 resolution (1280 x 720 per eye) and 30 FPS in a Wi-Fi6 connection \cite{b27}. A mini desktop with 16GB RAM, an i7-10710U CPU, and an Intel UHD Graphics was installed on the robot as the WebRTC server which was powered by a 100w power bank. The HMD and monitor were connected to a desktop with 16GB RAM, an i7-9700k CPU, a GeForce GTX 2080Ti dedicated GPU.

Participants used an Xbox wireless controller\footnote{Xbox wireless controller: \url{https://www.microsoft.com/en-us/p/xbox-wireless-controller/8xn59crbsqgz?cid=msft_web_collection.}} as the input device to control the UGV. Fig.~\ref{fig4}  (next page) shows the elements of the control and image transmission system consisting of a custom control protocol, robot controller software for Windows, Inter-Process-Communication (IPC), VR and PC interface built in Unity3D, and Xbox controller for controlling the robot remotely.

\subsection{Conditions}
In this study, the following four conditions were explored (see Fig.~\ref{fig1}). Fig.~\ref{fig1}e shows the UGV's camera pointing at a board with black dots and words to show the difference between monoscopic and stereoscopic images.

\begin{itemize}
    \item \textit{D1: Non-immersive display with fixed screen using monoscopic images} (see Fig.~\ref{fig1}A, 1a). We used a traditional 27-inch monitor as the display screen. The content of the display was from monoscopic images (fed from one of the binocular cameras placed on the robot). This is the non-immersive display and the baseline condition as this is what is provided typically in current UGV/UAV via a mobile phone, tablet or regular desktop.
    \item \textit{D2: Immersive display with fixed screen using monoscopic images} (see Fig.~\ref{fig1}B, 1b). We constructed a fixed virtual screen in VR. All irrelevant information is blacked out, except for what is displayed. This condition is the monoscopic view in VR. 
    \item \textit{D3: Immersive display with head tracking screen using monoscopic images} (see Fig.~\ref{fig1}C, 1c). We had this condition because this is the same as the ones that are used for current UGV or UGV that provide VR capabilities. We turned the fixed screen in into a moving screen that follows users’ head motion. It used monoscopic images captured from one of the binocular cameras.
    \item \textit{D4: Immersive display with head tracking screen using stereoscopic images} (see Fig.~\ref{fig1}d). We used the same moving screen as \textit{D3}, but the content of display was stereoscopic images captured using both binocular cameras. Each camera would provide images to the corresponding eye (see Fig.~\ref{fig2}b).
\end{itemize}

\subsection{Tasks and Procedures}\label{SCM}

To investigate the performance of the four different types of views and display methods, participants had to drive the robot through a customized maze (Fig.~\ref{fig3}b). We installed a high-definition video camera and 8 motion tracking VICON\footnote{Vicon system: \url{https://www.vicon.com/}} cameras to capture the movement of the robot along the maze and to detect any collisions with the objects in the maze.  

For each condition, participants had to drive the robotic car remotely and maneuver it through the maze as fast as possible but without hitting or colliding with the obstacles. The driving maze was designed to have five different tasks (see Fig.~\ref{fig3}a):

\begin{itemize}
\item \textit{T1}: This task consisted of 36 small and static wooden cubes. As the UGV moves deeper into the maze, the distance between the two cubes would become smaller (from 32cm to 30cm to 28cm).
\item \textit{T2}: This task had two big cubes placed horizontally. The user needs to drive UGV in a side-way manner to prevent collisions.
\item \textit{T3}: This task had a cardboard box that rotated at a constant speed. Participants would need to judge distance and time their move accordingly to avoid colliding with this spinning obstacle.
\item \textit{T4}: This task had a circular moving obstacle and a cross shaped, static obstacle. Participants had to judge the distance between the robot and the obstacles and time its move accordingly to avoid hitting both the moving obstacle and horizontal cross at the same time.
\item \textit{T5}: This task consisted of an obstacle that would move in a straight line in a backward and forward manner and a cube placed vertically. Participants needed to judge the robot’s distance and time its move to avoid colliding with either object.
\end{itemize}

A simple driving training outside the maze was given to the participants before they started the formal trials. The purpose of this training was to give participants the chance to become familiar with the controls, the HMD, and controlling the robot. After this training, the participants were asked to run the formal trials. Each participant had three trials in each display methods and the order of display methods was pre-determined by a 4 × 4 Latin square to reduce any learning effect.

\subsection{Participants}
Eight participants (5 males and 3 females, aged between 20-29, mean = 24.5) were recruited for this experiment. Data collected from the pre-experiment demographics questionnaire show that that they all declared to be healthy and did not have any health issues, physical and otherwise. They all had normal or corrected-to-normal vision and did not suffer from any known motion sickness issues in their normal daily activities. None of them had any experience driving a UGV using HMD in FPV. As such, it was the first time for all 8 participants to drive a remote a UGV using an HMD in FPV.

This experiment has been approved by the University Ethics Committee at Xi'an Jiaotong-Liverpool University.  

\subsection{Hypotheses}
Based on our review of the literature and experiment design, we formulated the following four hypotheses:

\begin{itemize}
    
    \item \textit{H$_1$}: \textit{D4} would lead to the best overall performance in distance perception; \textit{D1} would lead to the worse overall performance than \textit{D2} and \textit{D3}; 
    
    \item \textit{H$_2$}: \textit{D4} would lead to the best local performance in the complex tasks (\textit{T3}, \textit{T4}, and \textit{T5}); \textit{D1} would lead to the worse local performance than \textit{D2} and \textit{D3} in these three tasks (\textit{T3}, \textit{T4}, and \textit{T5}); 
    
    \item \textit{H$_3$}: \textit{D4} would significantly reduce user demands and would be the most popular display method; \textit{D1} would have more user demands and less user preferences than \textit{D2} and \textit{D3}.
    
\end{itemize}

\subsection{Results}

\subsubsection{Overall Performance}

All participants understood the nature of the tasks and all recorded data were valid. If there was a collision and it lasted less than 1 second, then it was considered as one collision only. If it lasted longer than 1 second, for each 1 second of collision time, we counted it as one collision. We recorded the number of collisions for each trial by checking frame by frame the high-definition videos from the camera and VICON tracking system. A Shapiro-Wilk test for normality was performed on each of measures separately for each condition and show that they followed a normal distribution.

Fig.~\ref{fig5}a shows the average number of collisions per condition. A repeated measures ANOVA with Greenhouse-Geisser correction showed that the mean of the number of collisions differed significantly between display types (\textit{F}(1.722, 12.054) = 10.691, \textit{p} \textless .05). A Bonferroni post-hoc test revealed that the number of collisions was significantly lower for \textit{D2} and \textit{D4} (\textit{p} \textless .05) compared to \textit{D1}. There was no significant difference between \textit{D2}, \textit{D3} and \textit{D4} (\textit{p} \textgreater .05).

 A repeated measures ANOVA with Greenhouse-Geisser correction found that there was no significant difference between conditons (\textit{F}(1.481, 10.365) = 1.327, \textit{p} \textgreater .05). 

\subsubsection{Local Performance}

\begin{figure*}[htbp]
\parbox{1\textwidth}{
\centering
\includegraphics[width=1\textwidth]{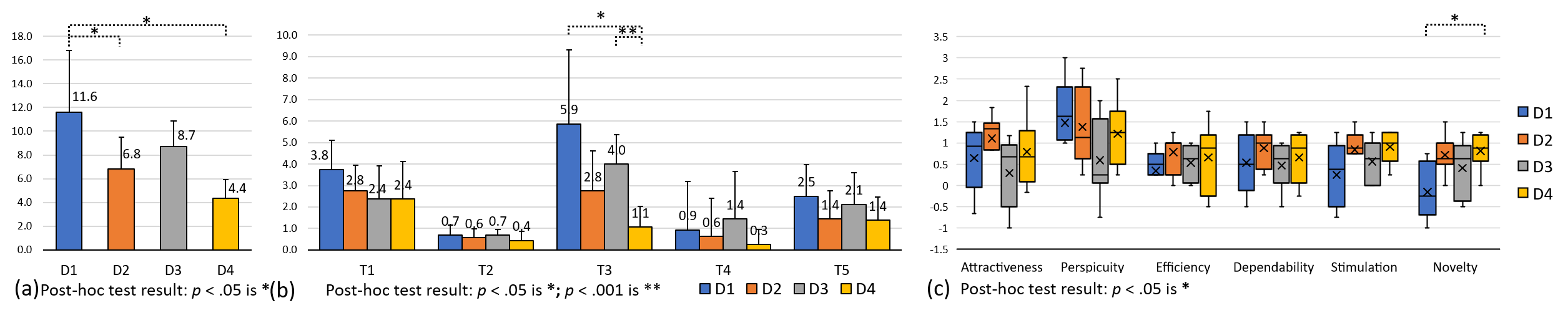}}
\caption{(a) Overall mean and std. deviation of number of collisions for the four conditions (\textit{D1}, \textit{D2}, \textit{D3}, and \textit{D4}); (b) Local mean and std. deviation of number of collisions for the four conditions (\textit{D1}, \textit{D2}, \textit{D3}, and \textit{D4}) in each task (\textit{T1}, \textit{T2}, \textit{T3}, \textit{T4}, \textit{T5}). All Error bars show +1 SD; (c) Average of response scores for each element of UEQ. Error bars show +1 SD. Higher scores represent more preferences in all cases. The '$\times$' symbol represents the mean value. Error bars show $\pm$1 SD.}
\label{fig5}
\end{figure*}

Fig.~\ref{fig5}b shows the average number of collisions for each condition in each task. A repeated measures ANOVA with Greenhouse-Geisser correction found that there was a significant difference in \textit{T1} (\textit{F}(1.464, 10.248) = 14.477, \textit{p} \textless .05). A Bonferroni post-hoc test showed that the number of collisions was significantly lower in \textit{T3} for \textit{D1} (\textit{p} \textless .05) and \textit{D3} (\textit{p} \textless .001) when compared to C4.

There was no significant difference among \textit{D1}, \textit{D2}, \textit{D3} and \textit{D4} in completion time.

\subsection{Subjective Results}

\subsubsection{NASA-TLX Workload}
Kruskal-Wallis H Test was conducted and found no significant difference in any sub-scales of the NASA-TLX workload (Mental, Physical, Temporal, Effort, Performance, and Frustration).

\subsubsection{User Experience Questionnaire (UEQ)}
Fig.~\ref{fig5}c shows the results of the UEQ. They were evaluated with the Kruskal-Wallis H Test comparing the four conditions for the six UEQ elements (Attractiveness, Perspicuity, Efficiency, Dependability, Stimulation, and Novelty). There was a significant difference in Novelty ($\chi^{2}$(3) = 9.077, \textit{p} \textless .05). Dunn’s post-hoc analysis showed a significant difference (\textit{p} \textless .05) for \textit{D1} vs \textit{D4} in Novelty.
\subsubsection{Interviews Results}
Overall, all participants had a positive experience in the experiment. None of the participants commented that they had any serious discomfort or simulation sickness in the four conditions. Most said that the most difficult task was \textit{T3}, the section with a cardboard box that rotated at a constant speed.
\subsection{Discussion}
The results confirmed some of our initial hypotheses but also revealed some different effects. We discuss these next.
\subsubsection{Overall Performance}
From the evaluation of the overall number of collisions, we have following effects:
\begin{itemize}
    \item \textit{D2} significantly reduced the number of collisions compared to \textit{D1}. This supports in part \textit{H$_{1}$}.
    \item \textit{D3} did not significantly reduce the number of collisions compared to \textit{D1}. This seems to contradict {\textit{H$_{1}$}}.
    \item \textit{D4} significantly reduced the number of collisions compared to \textit{D1}. This also supports in part {\textit{H$_{1}$}}.
\end{itemize}

\textit{D2}, \textit{D3}, and \textit{D4} all used VR. Their advantage over a normal monitor was that they could increase immersion and eliminate the reflection of light on the 2D display screen. Participants seem to have had better concentration and immersion when the off-screen information was all blackout. According to {\textit{H$_{1}$}}, all conditions with VR (\textit{D2}, \textit{D3}, and \textit{D4}) should have improved the performance compared to \textit{D1}. However, we only found significant improvements in \textit{D2} vs \textit{D1} and \textit{D4} vs \textit{D1}. This seems to indicate that the existing display modes from products in the market (\textit{D3} vs \textit{D1}), as mentioned in Section II.A (Condition), did not show significant difference in performance for real-time obstacle avoidance tasks. That is, they did not lead to a good performance overall.

For overall completion time, our analysis indicated that the different conditions did not significantly improve efficiency in the tasks, which contradicted \textit{H$_{1}$}.

In interviews after the experiment, participants commented that they needed more effort in \textit{T3} because estimating the distance of the UGV to a moving obstacle with frequent changes in  depth perception was very challenging.

\subsubsection{Local Performance}
Users behaved differently in different tasks in our experiment. Based on the analysis of collisions, we found that only one task (\textit{T3}) presented significant differences on performance, while the other tasks did not, which supported part of {\textit{H$_{2}$}} (only \textit{T3} showed differences). In particular, we found the following:
\begin{itemize}
    \item Using the virtual screen in VR (\textit{D2}) and monoscopic view in VR (\textit{D3}) did not significantly reduce the number of collisions compared to the 2D screen (\textit{D1}) in \textit{T3}.
    \item Having stereoscopic images in VR (\textit{D4}) significantly reduced the number of collisions compared to the 2D screen (\textit{D1}) and the monoscopic view in VE (\textit{D3}), respectively.
\end{itemize}

\textit{T3} required participants to judge the distance between the UGV and a spinning box. The difference between this task (\textit{T3}) and other tasks with dynamic obstacles (\textit{T4} and \textit{T5}) was that it not only changed the motion states of the obstacle but also significantly changed the depth of the obstacle that user needed to perceive from the views. This was where the advantages of 3DS view was observed the most because it allowed participants to obtain a better depth perception and 3D effect. Therefore, the participants could have a better sense of the distance between the UGV and the obstacles, which helped improve their performance. This was supported by the interview data. Participants commented that they needed more effort on \textit{T3} because estimating the distance to a moving obstacle with frequent changes in perceptual depth was quite challenging.

\subsubsection{User Demands and User Preferences}
The positive results from NASA-TLX data do not fully show equal preference of users for the four conditions, which contradicted the first part of {\textit{H$_{3}$}}. The results of UEQ gave us further insights into participants’ preferences. From the results of UEQ, \textit{D4} was considered more creative and innovative when compared to \textit{D1}, which supported the second part of {\textit{H$_{3}$}}. 

\section{User Study 2}
\begin{figure}[htbp]
\parbox{3.4in}{
\centering
\includegraphics[width=3.4in]{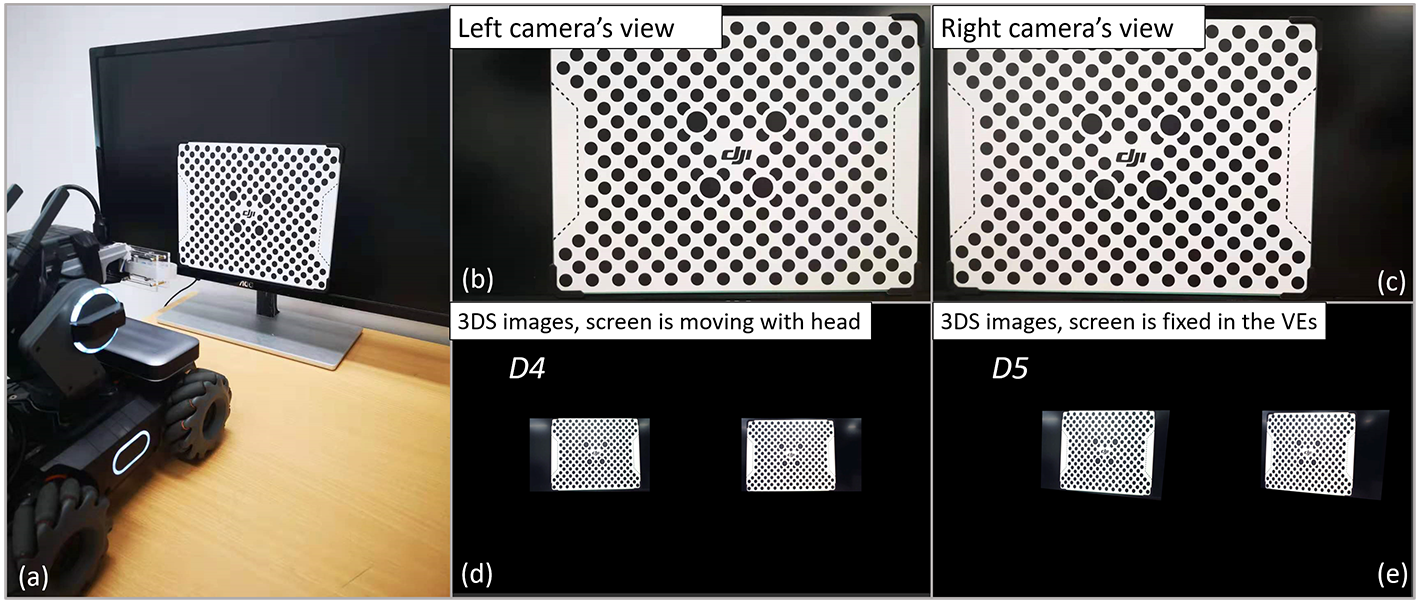}}
\caption{(a) A picture of the UGV's camera(s) watching a board. (b) View of the left camera of the binocular cameras. (c) View of the right camera of the binocular cameras. (d) Images for the two eyes in D4 and (e) in D5.}
\label{fig6}
\end{figure}

Based on the above results of Study 1, we combined \textit{D2} and \textit{D4} to create a new display mode - \textit{D5}. This version would show participants a fixed big screen in VR and allow them to have 3DS vision by giving each eye different images from the corresponding binocular camera. The only difference between \textit{D4} and \textit{D5} was whether the screen was movable or fixed (see Fig.~\ref{fig6}).

\begin{figure*}[htbp]
\parbox{1\textwidth}{
\centering
\includegraphics[width=1\textwidth]{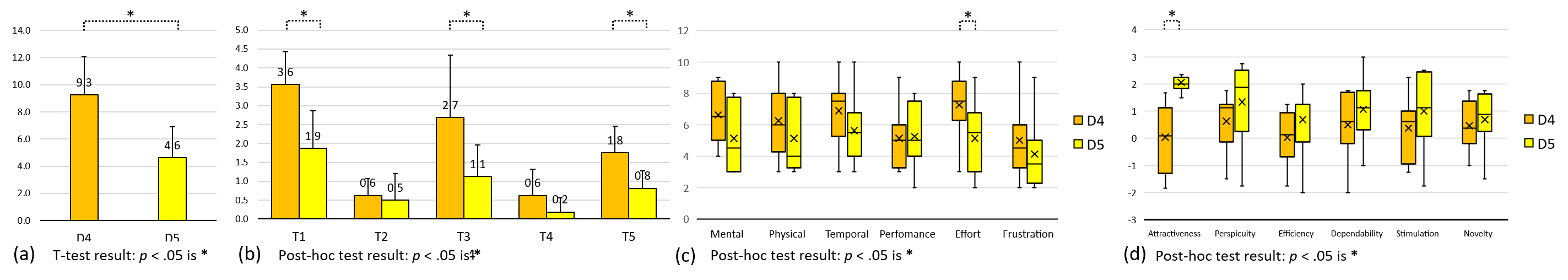}}
\caption{(a) Overall mean and std. deviation of number of collisions for the two conditions (\textit{D4} and \textit{D5}); (b) Local mean and std. deviation of number of collisions for the two conditions (\textit{D4} and \textit{D5}). All Error bars show +1 SD; (c) Average of response scores for each element of NASA-TLX workload. Higher scores represent more demands in all cases; (d) Average of response scores for each element of UEQ. Higher scores represent more preferences in all cases. The '$\times$' symbol represents the mean value. Error bars show $\pm$1 SD.}
\label{fig7}
\end{figure*}

\subsection{Conditions}
We decided to compare the best-performing version (\textit{D4}) in Study 1 with the new mode (\textit{D5}):
\begin{itemize}
    \item \textit{D4}: Immersive display with head tracking screen using stereoscopic images. The same as in Study 1.
    \item \textit{D5}: Immersive display with fixed screen using stereoscopic images. We used a fixed virtual screen in VR. The display content was stereoscopic (using both binocular cameras; each camera provides different images to their corresponding eye).
\end{itemize}

\subsection{Participants}
Another 8 participants (4 males and 4 females, aged between 21-27, mean = 22.5) were recruited for this experiment. Data from the pre-experiment questionnaire show that none of the participants had major physical discomfort, health problems, simulator sickness, or vision issues. All of them were able to complete the pre-training successfully. Similar to participants in Study 1, none of these participants had any experience with driving a UGV using HMD in FPV. As such, it was also the first time for all 8 participants to drive a UGV using an HMD in FPV.

\subsection{Experiment setup}
The rest of experiment setup was the same as in Study 1.
\subsection{Hypotheses}
Based on Study 1, we formulated the following two hypotheses:
\begin{itemize}
    \item \textit{H$_{4}$}: According to the objective results of Study 1, \textit{D5} would perform better than \textit{D4} overall and especially in \textit{T3}.
    \item \textit{H$_{5}$}: According to the subjective results of Study 1, \textit{D5} would be more preferred by participants than \textit{D4}.
\end{itemize}
\subsection{Results}
A Shapiro-Wilk test for normality was performed on each of measures separately for each condition and showed that they all followed a normal distribution. Fig.~\ref{fig7}a and 7b show the performance of the two modes. From these figures, we can see that:
\begin{itemize}
    \item For overall performance, the independent-samples t-test showed that \textit{D5} significantly reduced the number of collisions (\textit{t}(78) = 3.473, \textit{p} \textless .001) compared to \textit{D4}.
    \item In local performance, the t-test showed that \textit{D5} significantly reduced the number of collisions compared to \textit{D4} in \textit{T1} (\textit{t}(14)=3.473 = 3.631, \textit{p} \textless .05), \textit{T3} (\textit{t}(14) = 2.395, p \textless .05), and \textit{T5} (\textit{t}(14) = 3.147, p \textless .05).
\end{itemize}

Fig.~\ref{fig7}c and 7d show the summary of the NASA-TLX and UEQ questionnaire data. An analysis shows that: 
\begin{itemize}
    \item There was a significant difference in the NASA-TLX workload in Effort (\textit{$\chi^{2}$}(1) = 7.345, p \textless .05). Participants needed more effort using \textit{D4}.
    \item There was also a significant difference in the UEQ data in Attractiveness  (\textit{$\chi^{2}$}(1) = 10.678, p \textless .05). Participants preferred \textit{D5} significantly more.
\end{itemize}

\subsection{Discussion}
From the results, we can observe that participants performed better in \textit{D5}, as indicated by an overall lower number of collisions (especially in \textit{T3}). This result confirms {\textit{H$_{4}$}}. However, we also found that \textit{D5} had better performance in \textit{T1} and \textit{T5}, which we had not thought it would be the case. The NASA-TLX results also show a similar trend. Participants did not need much effort to finish the tasks using \textit{D5}. In addition, the UEQ data showed that participants significantly liked \textit{D5} more, which supported {\textit{H$_{5}$}}. All the above results show that \textit{D5} led to better performance than \textit{D4} in overall (local) performance, required lower demands from participants, and led to an enhanced user experience. These results suggest that an immersive display with a fixed screen using stereoscopic images (like \textit{D5}) is a viable approach to allow operators remotely control a UGV when obstacle avoidance is essential.

\section{Limitations and Future Work}
From the results of the two studies, we can observe that an immersive display with stereoscopic images can significantly increase user performance in obstacle avoidance tasks. They also indicate that a fixed screen in VR led to a better performance than a screen that moves together with head motions.

This research has the following two limitations, which can serve as directions for future work. Even with binocular cameras, it is still not possible to significantly increase the environmental information because of the limitations of the camera's FOV. If users do not manipulate the camera, they are not able to see a wide view of the environment on a single screen. Cameras that can capture wide views, including 360° panoramic ones, around the robot could be one possible solution but further research is needed to assess their suitability because such views will require users to move their heads frequently, which could lead to higher levels of motion sickness.

In this research, we explored views on two types of displays and did not investigate image distortion or enhancement approaches. Given that the immersive VR display with stereoscopic view led to better performance and higher usability, it will be interesting to see if distorting the images (to enhance certain elements for example) could allow even better performance. Applying edge enhancement could potentially improve the perception of the contours of obstacles and may let operators gauge distance information in a more precise way. This line of research could produce useful and interesting results and applications.

In addition, another way to help improve manoeuvrability to avoid obstacles is to use operators' physiological information that can be captured during teleoperation. For example, eye gaze is readily available and implicitly provided by operators. Recent research shows that gaze data can be used to improve object manipulation in VR environments \cite{b40}. Typically, when approaching an obstacle to be avoided, operators' gaze would likely be fixated on that object and, when a fixation is longer than a threshold, the system can provide additional information to help steer the UGV towards the most optimal path. As eye trackers are becoming a common feature of HMD, using gaze data could represent a low-cost and efficient approach for improving operator-UGV interaction. Further research is needed to explore how this can be achieved.

\section{Conclusion}
In this paper, we have explored four viewing/display modes for real-time unmanned ground vehicle (UGV) control in obstacle avoidance tasks. The aim is to investigate whether these modes allow users to determine UGV-to-obstacle distances. Study 1 evaluated the performance and user preference of the four modes in a maze with moving and static objects that the UGV had to avoid. Results from this study show that the version with stereoscopic images displayed in a virtual reality head-mounted display (VR HMD) led to better performance and usability. In Study 2, a new display mode that combines two modes from Study 1 was compared with the best-performing mode (VR with stereoscopic view). Overall, our results indicate that an immersive VR display with a fixed screen using stereoscopic images is an applicable and suitable approach for improving depth perception when controlling a UGV in real-time in obstacle avoidance tasks, whether static or moving. It also helped participants lower their workload levels and led to an enhanced user experience. 

\section*{Acknowledgment}
The authors would like to thank the participants for their time and the reviewers for their reviews and useful comments. 
\bibliographystyle{./bibliography/IEEEtran}
\bibliography{./bibliography/Reference}

\vspace{12pt}
\end{document}